\documentclass[proceedings]{rmaa}
\usepackage{rmaacite}

\renewcommand{\P}[1]{%
\ifnum#1=1\hbox{OW~168--326E}\fi
\ifnum#1=2\hbox{OW~167--317}\fi
\ifnum#1=3\hbox{OW~163--317}\fi
\ifnum#1=5\hbox{OW~158--323}\fi
\ifnum#1=0\hbox{OW~171--334}\fi}

\title{Photometric Properties Of Low-Redshift Galaxy Clusters}
\author{Omar L\'opez-Cruz\altaffilmark{1} \affil{INAOE-Tonantzintla
\& Department of Astronomy, University of Toronto} }
\altaffiltext{1}{Visiting Astronomer, Kitt Peak National Observatory,
National Optical Astronomy Observatories, which is operated by the
Association of Universities for Research in Astronomy, Inc (AURA)
under cooperative agreement with the NSF.}

\fulladdresses{
\item Omar L\'opez-Cruz: Instituto Nacional de Astrof{\'\i}sica,
Optica y Electr\'onica (INAOE-Tonantzintla). Luis Enrique Erro No.1,
Tonantzintla, Puebla, Puebla 72840, M\'exico, (omar@inaoep.mx)}

\shortauthor{L\'opez-Cruz}
\shorttitle{Photometric Properties of Low-Redshift Galaxy Clusters}


\abstract{A comprehensive multicolor survey was undertaken to
investigate global optical properties of Abell clusters of
galaxies. This survey was christened the {\em Low-Redshift Cluster
Optical Survey} (LOCOS). LOCOS was devised to search for patterns of
galaxy evolution induced by the environment.  The generated data base
contains accurate deep CCD photometric measurements (Kron-Cousins
$R,\,B ~{\rm and}~ I$) for a sample of 46 low-redshift ($0.04 \leq z
\leq 0.18$) Abell clusters. This is one of the few large surveys that
included the contribution due to dwarf galaxies (about 5.5 mag deeper
than the $R$ characteristic magnitude (M$_{R}^{*}$); ${\rm
Ho}=50\,h_{50} km\,s^{-1}\,Mpc^{-1}$, ${\rm qo}=0$). Due to space
restrictions only the main results concerning the variations at the
bright-end of the luminosity function (LF) are presented here. Other
results are presented elsewhere (L\'opez-Cruz \& Yee 2000a,b).  We
have detected clear variations at both the bright end and the faint
end of the LF. The nature of the variations at the bright end revealed
that poor cD clusters have dimmer M$_{R}^{*}$. We can explain these
variations as a result of dynamical friction. On the other hand,
non-cD clusters seem to have unaffected LFs. A third class termed as
binary clusters seems to be a transition class that might have
resulted from cluster-cluster mergers.}

\resumen{El levantamiento multicolor para c\'umulos de galaxias
cercanos LOCOS se realiz\'o para investigar las propiedades generales
de los c\'umulos y la busqueda de patrones de evoluci\'on gal\'actica
inducida por el medio ambiente. LOCOS incluye la observaci\'on de 46
c\'umulos cercanos del cat\'alogo de Abell en las bandas $R, B,$ e $I$
del sistema fotom\'etrico Kron-Cousins. LOCOS es uno de pocos
levantamientos extensos, donde se ha incluido la contribuci\'on de las
galaxias enanas. En esta contribuci\'on s\'olo se presentan los
resultados concernientes a las variaciones de la Funci\'on de
Luminosidad (FdeL), otros resultados ser\'an presentados en
L\'opez-Cruz \& Yee (2000a,b). Hemos detectado variaciones en ambas
ramas de la FdeL. Dichas variaciones no son totalmente
err\'aticas. Por ejemplo, al investigar la naturaleza de las
variaciones en la rama brillante de la FdeL se encontr\'o que la
magnitud carater{\'\i}stica M$_{R}^{*}$ para cumulos cD pobres es
m\'as d\'ebil que la de los c\'umulos cD ricos. Los c\'umulos que no
tienen galaxias cD parecen no afectados y presentan la misma
M$_{R}^{*}$ dentro de los margenes de error. Mientras que los
c\'umulos binarios muestra ser una clase en transici\'on al ser,
posiblemente, el resultado de la fusi\'on de dos  c\'umulos.}
\nonstopmode

\begin{document}

\maketitle

\section{Introduction}
\label{sec:intro}

It is distressing that, up to now, we do not have a clear idea about
the true shape of the LF for galaxies in the field or in
clusters. Disagreement has existed ever since the earliest attempts to
determine the galaxy LF. Hubble in the 1930's advocated a Gaussian
shape LF for bright galaxies, while Zwicky in the late 1950's
suggested an exponential increase with decreasing magnitude.  It took
accurate photometric measurement to show that both descriptions were
right. The shape of the LF has a strong dependency on galaxy type and
luminosity \cite{BST88}: a Gaussian function fits the LF of giant
galaxies and Zwicky's exponential law fits the LF of dwarf
galaxies. One the same grounds the subject of the universality of the
LF has always been questioned. Universality was first suggested by
\scite{Ab62} and reintroduced, in an analytical manner, by
\scite{Sch76}. Many photographic and, very few, digital studies have
followed since then, but the subject has been far from being
settled. For example, no universality was first suggested in the
pioneering work of \scite{Oe74}. However the most recent work of
\scite{Lum97} indicates otherwise. One is inclined to think that under
the influence of dynamical effects the LF ought to be non-universal
\cite{Dre84}. Moreover if the mixture of galaxies varies from cluster
to cluster, then the universality of the cluster galaxy LF is not expected
\cite{BST88}. The LOCOS sample contains rich Abell clusters that have
strong X-ray emission \cite{JF99}. It is one the largest digital
cluster surveys to date. The results that are presented below suggest
that the non-universality of the LF is due to changes
induced by the environment.

\begin{figure}[b]
  \begin{center}
    \leavevmode
    \includegraphics[height=9 cm,width=5 cm,angle=90]{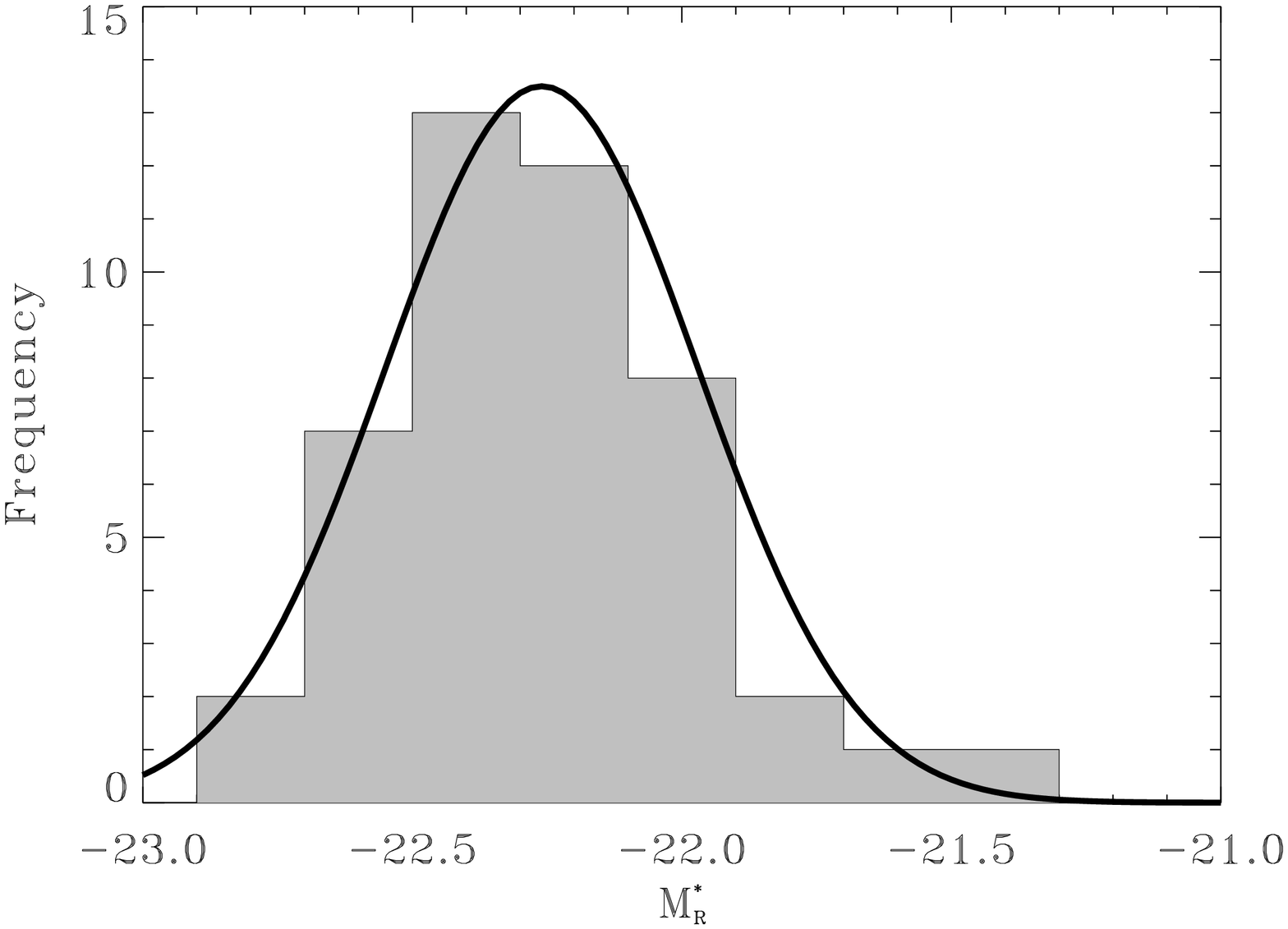}
    \caption{The number distribution of
M$_{R}^{*}$. A Gaussian distribution with $<{\rm M}_{R}^{*}> =-22.26$ mag
and dispersion $\tilde{\sigma} =0.29$ mag describes the distribution.}
    \label{fig:gauss}
  \end{center}
\end{figure}

\section{Observations and Luminosity Function Generation}
\label{sec:observations}
The observations were carried out at KPNO with the 0.9m telescope and
the T2KA CCD ($2048\,\times\,2048$) pixels in Kron-Cousins $I$, $R$
and $B$ bands with  a field of view of
$23.2\arcmin\,\times\,23.2\arcmin$ with a scale of $0.68\arcsec/{\rm
pixel}$. The integration times varied from 900 to 2500 seconds,
depending on the filter and the redshift of the cluster. The
photometric calibrations were done using stars from
\scite{La92}. Control fields are also an integral part of this survey;
we observed 5 control fields (in $R$ and $B$) chosen at random
positions in the sky about $5\deg$ away from a cluster observation.
Data preprocessing was done with IRAF, while the object finding,
star/galaxy classification, photometry, and the generation of
catalogues was done with {\sf PPP} \cite{YEC96}. More details are
presented in \cite{LCYa00}.

The LF functions were generated by background subtraction using the
counts from the control fields. The main differences with respect to
other studies are that all the photometry and calibrations were based
on CCD observations and that color information was used to generate
the background subtracted-LF, i.e., galaxies whose colors were too red
with respected to the color-magnitude relation were rejected by
applying color-cuts in the color-magnitude space. In general, two
Schechter functions were used to fit the counts using a $\chi^{2}$
minimization (see \pcite{LCY95}). The entire cluster sample was
combined and a Schechter fit to the bright-end (galaxies with
M$_R\,\le\,-20$ mag) of the combined LF gave an slope
$\alpha_{com}=-1.04\pm0.05$ and a characteristic magnitude
M$^{*}_{com}=-22.53\pm0.09 + 5\log\,h_{50}$ mag (cf. \pcite{Ga97}). We
proceed to fit the bright-end of the LF of individual clusters with a
fixed $\alpha=-1.0$ and the constraint that the total number of
galaxies in the resulting fits was equal to the actual total number in
the observed LFs (e.g., \pcite{Dre78,Co89}).  Full details regarding
the generation of the LFs are given in \scite{LC97} and
\scite{LCYb00}.

\section{LF Variations at the   Bright End}
\label{sec:variations}

Since the effect of contamination by background clusters is  small at
bright magnitudes, the entire cluster sample can be used to search for
variations of M$_{R}^{*}$.  The distribution of M$_{R}^{*}$ is shown
in Figure 1.  It is found that the values of M$_{R}^{*}$ are normally
distributed with mean, $<{\rm M}_{R}^{*}> = -22.26$, and standard
deviation, $\tilde{\sigma} = 0.29$.  Our $<{\rm M}_{R}^{*}> $ agrees
to within 0.2 mag with the mean values previously reported (e.g.,
\pcite{Sch76,Dre78,Co89}), although a ``canonical'' $\alpha=-1.25$ was
used in their fittings. In Figure 1 we note that there is a slight
tail towards faint magnitudes in the distribution.  The average
error of M$_{R}^{*}$ for the whole sample is $0.21\pm0.07$ mag.
Hence, the dispersion in the distribution of M$_{R}^{*}$ is somewhat
larger than that expected from the uncertainties of the measurements,
indicating some chance of departure from universality.

\begin{figure}[h]
  \begin{center} \leavevmode \includegraphics[height=9 cm,width=5
    cm,angle=90]{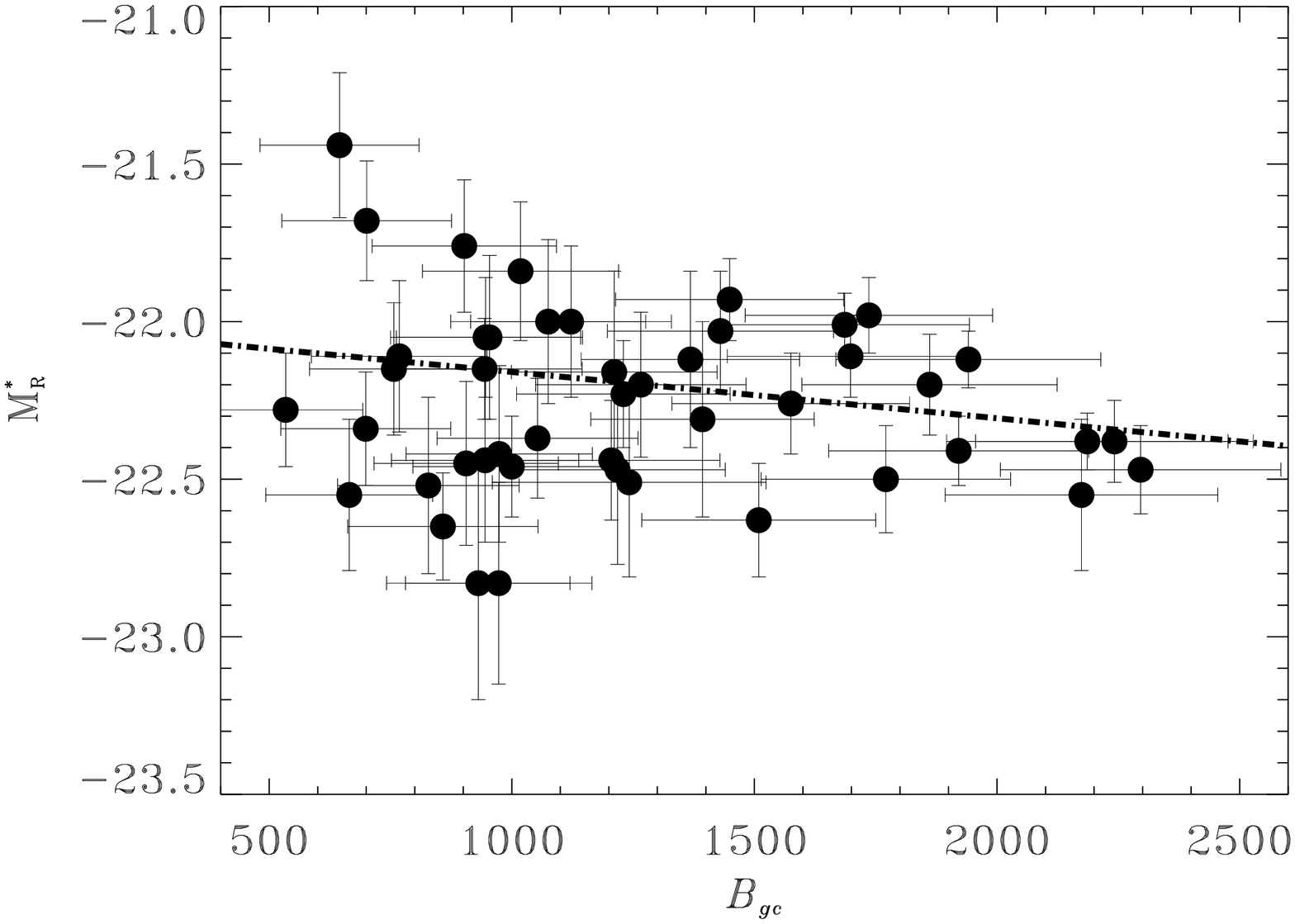} \caption{The variation of the
    M$_{R}^{*}$ with richness ($B_{gc}$) for the whole sample. There
    is a weak correlation between M$_{R}^*$ and $B_{gc}$. }
    \label{fig:mvsbgc} \end{center}
\end{figure}

In order to investigate whether these variations are associated with
intrinsic cluster properties, we compare the values of M$_{R}^{*}$ and
the richness indicator $B_{gc}$ \cite{YLC99}.  Figure 2 shows the
variations of M$_{R}^{*}$ with richness.  There is a marginal
correlation between M$_{R}^{*}$ and $B_{gc}$ (significant at the 15\%
level). To investigate this possible correlation further, we divide
our sample into three groups based on the cluster morphological
properties.  The first group has objects that have a very strong
indication that their brightest cluster members (BCMs) are cD
galaxies\footnote{We call cD galaxy a BCM that has the presence of an
extended luminous halo, we also recognize two classes: halo, according
to Schombert's (1988) criterion for cD-ness, and halo-dominated cDs,
as in \scite{LCE97}.}. These clusters are Rood-Sastry type ``cD'' and
have Bautz-Morgan types between I and I-II.  For the second group we
select all those objects with Rood-Sastry Type ``B''.  The reason for
selecting these objects as a separate class is because binary clusters
have indications of recent cluster-cluster merging \cite{Tre90}.  The
third group is defined, by simply negating the criteria that lead us
to select the cD and B clusters, i.e., clusters with Rood-Sastry type
different from cD or B, and Bautz-Morgan type later than I-II.  We
term these clusters non-cD clusters. If cD galaxies are formed by a
dynamical evolution mechanism, then non-cD clusters should have
properties closer to unevolved clusters. The classes cD and non-cD are
well separated, see below. However, we believe that B clusters are a
mixture of cD and non-cD clusters.  If one or more of the binaries are
cD, then we would expect the cluster to have properties similar to cD
clusters. But the general properties will be more difficult to
characterize because of the recent cluster-cluster merger effects.

The table below summarizes our group definitions. There are five
clusters that do not belong to any of the defined groups: three with
ambiguous indication that their BCGs are cDs, A415 (BM II, R-S cD),
A665 (B-M III, R-S cD), A1413 (B-M I, R-S C), and two with incomplete
morphological information (A1569, A2555).
\vspace{0.5 cm}

\centerline{
\begin{tabular}{lccc}
Group &Bautz-Morgan Type&Rood-Sastry Type &\# of clusters\\\hline
cD&I, I-II&cD&12\\
non-cD&II, II-III, III& C, F, L,I&23\\
B & &B&6\\
unclassified & & &5\\
\end{tabular}
}
\vspace{0.5 cm}

\begin{figure}[h]
 \begin{center} \leavevmode \includegraphics[height=9 cm,width=5
    cm,angle=90]{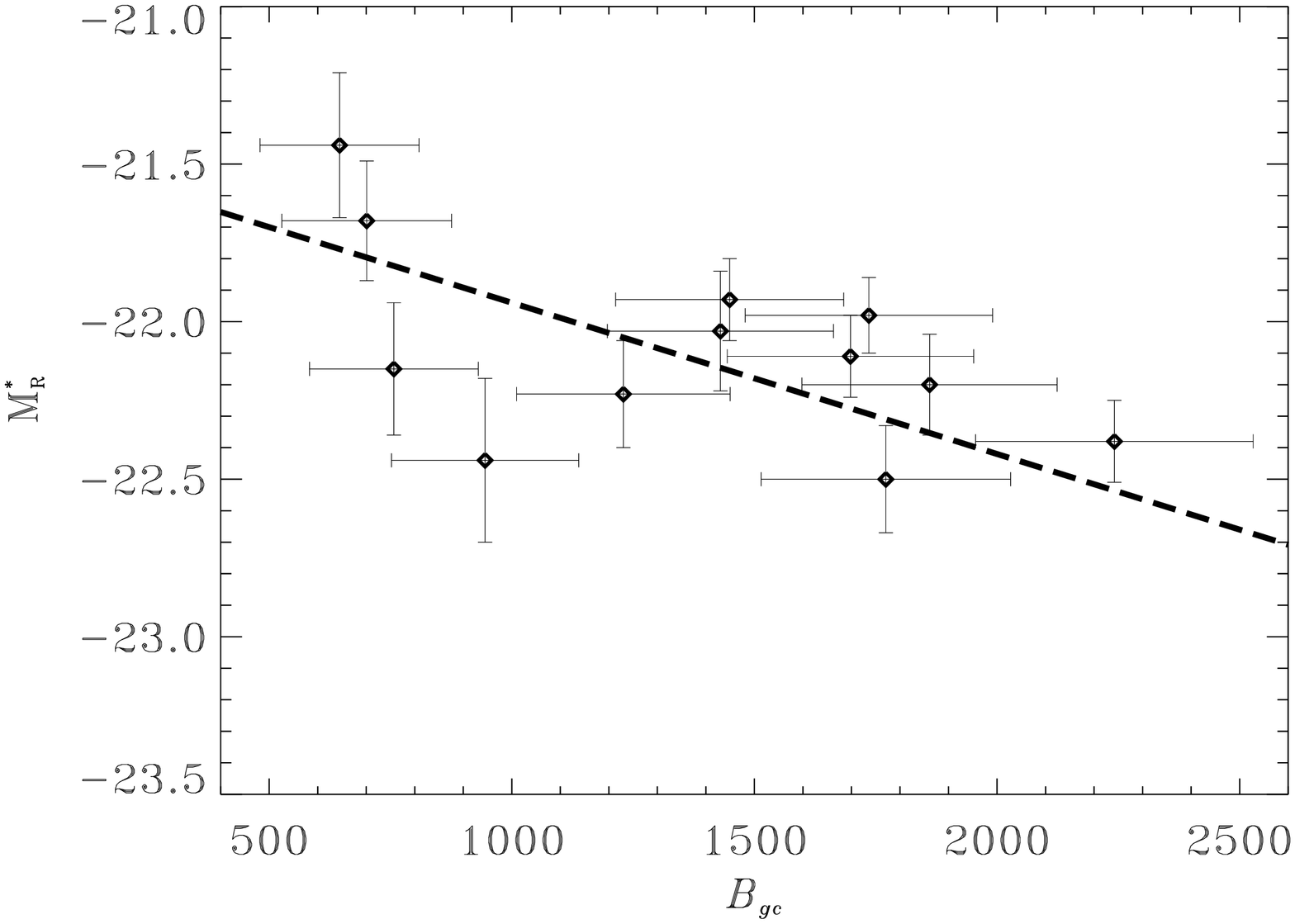} \caption{ A significant correlation
    between M$_{R}^{*}$ and $B_{gc}$ is found for cD clusters.}
    \label{fig:cD}
\end{center}
\end{figure}
\subsection{cD Clusters}
Figure 3 is a plot of M$_{R}^{*}~vs~B_{gc}$ for cD clusters. It shows
that there is a clear trend indicating that M$_{R}^{*}$ becomes
brighter with richness.  We can try to characterize this variation
with a linear fit. A least squares fit considering the errors in both
variables gives the following:
\begin{equation}
{M_{R}^*} = -21.46 (\pm 0.20) - 0.428(\pm 0.138)
\left(\frac{B_{gc}}{1\times10^{3}}\right).
\end{equation}
The correlation coefficient in Equation 1 is $r= -0.55$. For 12
points, the probability of exceeding $|r|$ in a random sample is
$P(r,\nu)=0.07$.  Therefore, we conclude that the correlation is
significant.  We suggest that Equation 1 is consistent with the the
galactic cannibalism model, in the sense that poorer cD clusters are
more effective in depleting their giant galaxies to form a cD
galaxy. From the dependence of the time scale for dynamical friction
with cluster velocity dispersion $\tau_{df}\propto\sigma_{cl}^3$ (if
the cluster potential is a singular isothermal well), and since
$\sigma_{cl}\propto0.4B_{gc}$ \cite{YLC99}. It follows that the
orbital decay rate is relatively faster in poor clusters than in the
rich ones.  Detailed calculations by \scite{Me88} and \cite{Tre90}
show that the decay rates due to dynamical friction are not short
enough to produce the total luminosity of cD galaxies (central galaxy
+ envelope). Equation 1 is consistent with an overall slow rate of
orbital decay, because if one or two giant galaxies are cannibalized
in a poor cluster, the change in M$_{R}^{*}$ is easily detectable.
Nevertheless, Figure 3 provides a case for a consequence of cannibalism
that is very suggestive and significant. Indeed, the correlation
depicted in Figure 3 does not disappear even when the four poorest
clusters are removed from the analysis. A more complete picture for cD
galaxy formation also involves the disruption of dwarf galaxies
\cite{LCE97}.We note that \scite{Dre78} suggested a converse trend.
He indicated that a fading of M$_{R}^{*}$ with richness was consistent
with his data.  However, his sample included only five Bautz-Morgan
clusters, four of which are part of our cD cluster group (A401, A1413,
A2029, A2670).  The fifth one is A2218 a Rood-Sastry C cluster.
However, Dressler's richness scale, which  was based on the central
counts scaled by the volume of the cluster core, and ours do not
agree, particularly for the cases of A2670, A401 and A1413. The main
source of uncertainty in Dressler's richness estimator is the size of
the clusters' core. If we use our richness scale, the trend depicted in
Dressler's Figure 5 disappears. We should remark that other studies
based on photographic material have only reported marginal cases of
this kind of variations (e.g., \pcite{Co89,Lum97}).

\subsection{Non-cD Clusters}

\begin{figure}
\begin{center}
    \leavevmode \includegraphics[height=9 cm,width=5
  cm,angle=90]{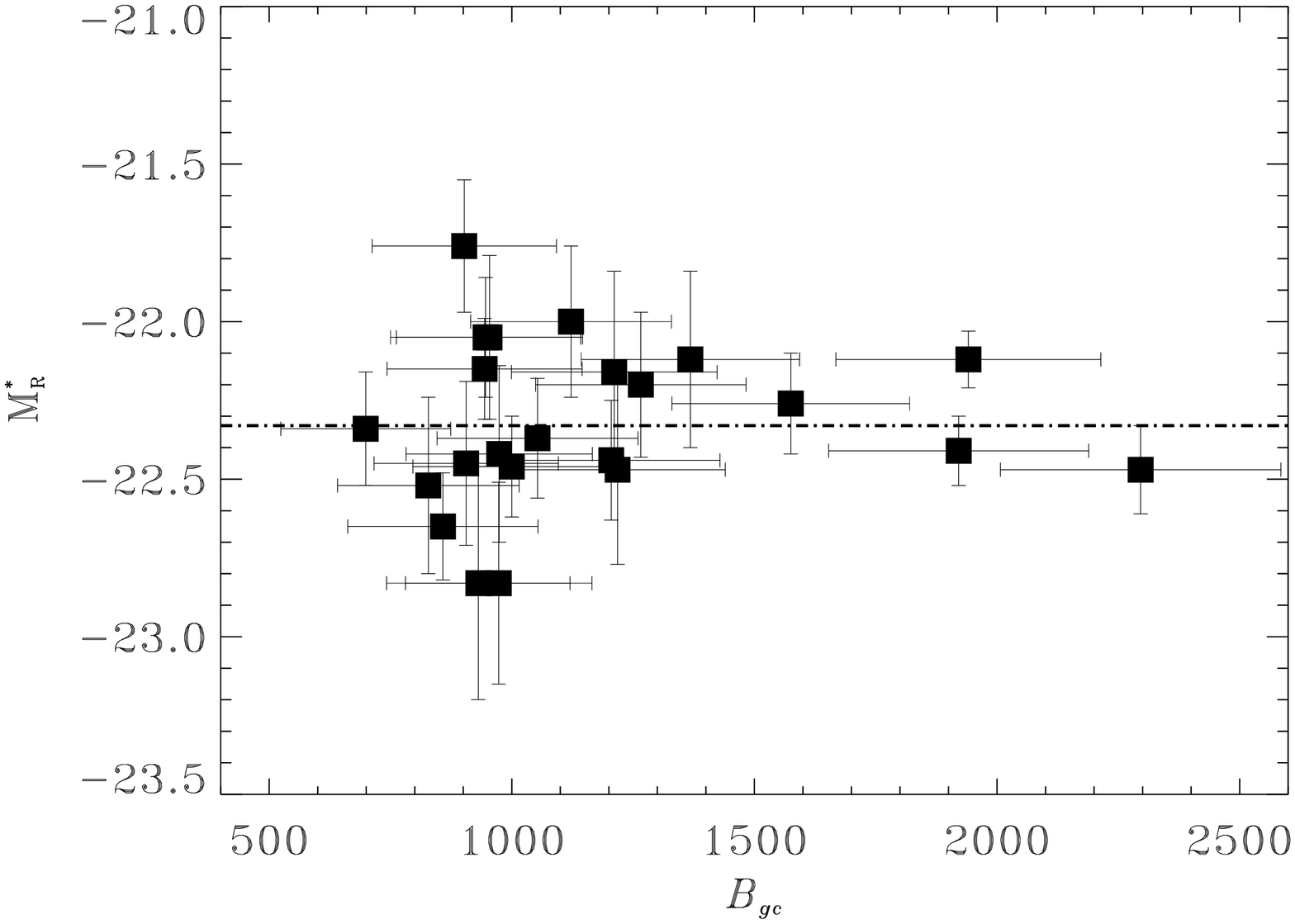} \caption{M$_{R}^{*}$ is uniformly
  distributed with richness ($B_{gc}$)} \label{fig:non-cD}
\end{center}
\end{figure}

Apart from the morphological differences between cD clusters and
non-cD clusters, in general it is found that non-cD clusters are
poorer than cD clusters. Moreover, we found that the non-cD clusters
show no correlation with $B_{gc}$ as depicted in Figure 4.  Hence, the
marginal correlation we see in Figure 1 arises entirely from the cD
clusters contribution. Non-cD clusters are normally distributed
with $<{\rm M}_{R}^{*}> =-22.33\,$mag, and $\tilde{\sigma}= 0.26\,$mag, which
is smaller than that from the whole sample. Hence, we conclude that
non-cD clusters are consistent with having a universal M$_{R}^{*}$
within the uncertainties of the observations.  Furthermore, a
Kolmogorov-Smirnov test shows that the distributions of M$^*$ and
$B_{gc}$ for the cD clusters and non-cD clusters are significantly
different ($D=0.36$, $p=0.21$; $D=0.41$, p=0.10, respectively). This
further, confirms our assertion that these  classes are well separated.
Most importantly, we also conclude that cD and non-cD clusters must
have different dynamical potentials. Therefore, their dynamical
evolution should be different. Poor clusters might represent regions
that turned around later than cD clusters.

\subsection{Binary Clusters}

Our idea that binary clusters are an intermediate class is fair. For
instance: in the Coma cluster there are three very luminous giant
galaxies. Two of those, NGC 4874 and NGC 4839 are halo-cDs. High
resolution X-ray images have shown two components of hot gas
associated with the galaxies NGC 4874 and NGC 4889,
respectively. Based on the galaxy distribution and the X-ray
distribution, it was concluded that NGC 4874 and NGC 4889 are the BCGs
of two clusters in the process of merging. Indeed, we found that B
clusters seem to behave differently from both cD or non-cD
clusters. M$_{R}^{*}$ fojr binary clusters is consistent with a uniform
distribution with $B_{gc}$ ($<{\rm M}_{R}^{*}>= -22.38$), but binary
clusters are richer than non-cD clusters on the average. However, with
such a small sample of binary clusters,  these results should not be
overemphasized.

\section{SUMMARY}

We found that the cluster LF is in general {\em not universal}. As
\scite{KD89}, we also suggest that cD galaxies are formed by special
processes.  The correlation between M$_{R}^{*}$ and $B_{gc}$ suggests
that galactic cannibalism contributes in the process of cD galaxy
formation. Indeed, if cD galaxies have grown by dynamical processes,
we expect a dimming of M$_{R}^{*}$ for earlier Bautz-Morgan type;
Figure 5 depicts such a trend. The same trend was suggested by
\scite{Trv96}. In contrast, we found that the bright end of LF for
non-cD cluster is universal and, hence, there is no dependence on the
environment. Therefore their properties are closer to that of the
field. Finally, we remark that B clusters seem to have properties
somewhat intermediate between cD and non-cD clusters. From the results
presented here, we propose that the characteristic magnitude may serve as a
good distance indicator once the effects of the environment are taken
into consideration.

\begin{figure}
\begin{center}
\leavevmode 
\includegraphics[height=9 cm,width=5 cm,angle=90]{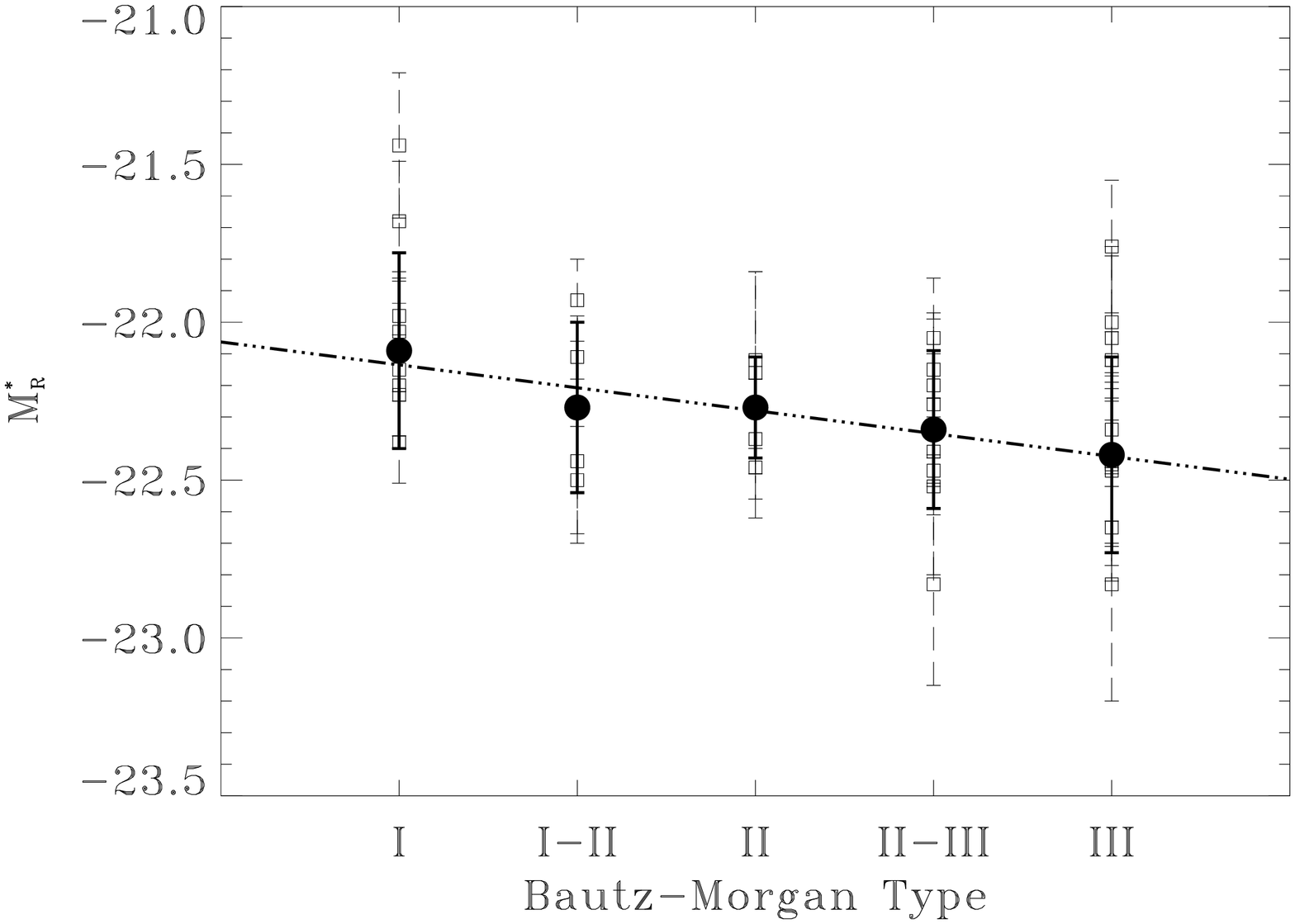}
\caption{A significant correlation is found between M$_{R}^{*}$ and
the Bautz-Morgan type for cD and non cD clusters. M$_{R}^{*}$ is about
half a magnitude dimmer for type I than for type III clusters. The
squares are the individual clusters. The filled circles are the median
M$_{R}^{*}$ for each B-M type .}
\label{fig:bm}
\end{center}
\end{figure}
\section{Acknowledgments}

The author is grateful to Prof. Howard K.C Yee for all these years of
collaboration and mentorship, to CONACyT-M\'exico (Beca de Postgrado,
Repatriaci\'on y Proyecto de Investigaci\'on Inicial),
INAOE, the Department of Astronomy and the SGS of the
UofT for sponsorship. OLC is also thankful to the SOC of the IX RRLA
of the IAU/UAI for the kind distinction conferred to his Ph.D. thesis, and 
W. Wall  for his comments.



\begin{thebibliography}

\bibitem[Abell<1962>]{Ab62} 
Abell, G.O. 1962, in {\em Problems of
Extra-Galactic Research}, G.C. McVittie ed., New York, Macmillan,
pp. 213-238

\bibitem[Binggeli et al.{}<1988>]{BST88}
{Binggeli}, B., {Sandage}, A., \& {Tammann}, G. A. 1988, ARAA, {26}, 509.

\bibitem[Colless<1989>]{Co89}
{Colless}, M. 1989, MNRAS, {237}, 799.


\bibitem[Dressler<1978>]{Dre78}
{Dressler}, A. 1978, ApJ, {223}, 765.

\bibitem[Dressler<1984>]{Dre84}
{Dressler}, A. 1984, ARAA, {22}, 185.

\bibitem[Gaidos<1997>]{Ga97} 
{Gaidos}, E.J. 1997, AJ, {113}, 117.

\bibitem[Jones \& Forman<1999>]{JF99}
{Jones}, C. \&  {Forman}, W. 1999, ApJ, {511}, 65.

\bibitem[Kormendy \& Djorgovski<1989>]{KD89}
{Kormendy}, J. \& {Djorgovski}, S.G. 1989, ARAA, {27}, 235.

\bibitem[Landolt<1992>]{La92}
{Landolt}, A. U. 1992, AJ, {104}, 372

\bibitem[L\'opez-Cruz \& Yee<1995>]{LCY95}
{L\'opez-Cruz}, O., \&  Yee H.K.C. 1995, A.S.P. Conf. Ser. {86}, 279.

\bibitem[L\'opez-Cruz<1997>]{LC97}
L\'opez-Cruz, O. 1997, Ph.D. Thesis, University of Toronto.

\bibitem[L\'opez-Cruz et al.{}<1997>]{LCE97} 
{L\'opez-Cruz}, Yee, H.K.C., Brown, J.P., Jones, C. \& Forman, W. 1997,ApJ,{476}, L97.

\bibitem[L\'opez-Cruz \& Yee<2000a>]{LCYa00}
L\'opez-Cruz, O., \& Yee H.K.C. 2000a, to be submitted to Rev. Mex. A \& A

\bibitem[L\'opez-Cruz \& Yee<2000b>]{LCYb00}
L\'opez-Cruz, O., \& Yee H.K.C. 2000b, to be submitted to ApJ.

\bibitem[Lumsden et al.{}<1997>]{Lum97} 
{Lumsden}, S. L. et al. 1997, MNRAS, {290}, 119.

\bibitem[Merrit<1988>]{Me88}
{Merritt}, D. 1988, A.S.P. Conf. Ser., {5}, 175. 

\bibitem[Oemler<1974>]{Oe74}
Oemler, A. Jr. 1974, ApJ, {194}, 1.

\bibitem[Schechter<1976>]{Sch76}
Schechter P. L.  1976, ApJ, {203}, 297.

\bibitem[Schombert<1988>]{Sch88}
{Schombert}, J.M. 1988, ApJ, {328}, 475

\bibitem[Tremaine<1990>]{Tre90}
{Tremaine}, S. 1990,~in {\em Dynamics and Interactions of Galaxies},
R. Wielen, ed., Heidelberg, Springer Verlag, pp. 394-405

\bibitem[Tr\`evese et al.{}<1996>]{Trv96}
Tr\`evese, D., Cirimele, G. \& Appodia, B. 1996, A\&A, {315}, 365. 

\bibitem[Yee, et al.{}<1996>]{YEC96}
{Yee}, H.K.C., {Ellingson}, E., \&  {Carlberg}, R.G. ApJS, 1996, {102}, 269.

\bibitem[Yee \& L\'opez-Cruz<1999>]{YLC99}
Yee, H.K.C. \& L\'opez-Cruz, O. 1999, AJ, {117}, 1985.

\end{thebibliography}
\end{document}